\documentclass[letterpaper,twocolumn,english,aps,prb,floatfix,showpacs,amsfonts]{revtex4}
\usepackage[T1]{fontenc}
\usepackage[latin1]{inputenc}
\usepackage{amsmath}
\usepackage{graphicx}
\usepackage{amssymb}

\makeatletter

\providecommand{\LyX}{L\kern-.1667em\lower.25em\hbox{Y}\kern-.125emX\@}

\usepackage{bm}

\newcommand{\bt} { \begin{tabular} }
\newcommand{\et}{ \end{tabular} }
\newcommand{\bc} { \begin{center} }
\newcommand{\ec}{ \end{center} }
\newcommand{\mc}{ \multicolumn }

\newcommand{\bfi}{\begin{figure} }
\newcommand{\efi}{\end{figure} }

\newcommand{\btb} { \begin{table} }
\newcommand{\etb}{ \end{table} }

\usepackage{babel}
\makeatother
\begin{document}

\title{The spin and charge gaps of the half-filled N-leg Kondo ladders }

\author{J. C. Xavier}

\affiliation{Instituto de F\'{\i}sica Gleb Wataghin, Unicamp, C.P. 6165, 13083-970 Campinas
SP, Brazil}

\date{\today{}}

\begin{abstract}
In this work, we study N-leg Kondo ladders at half-filling through the density matrix
renormalization group. We found non-zero spin and charge gaps for any finite number
of legs and Kondo coupling $J>0$. We also show evidence of the existence of a quantum critical
point in the two dimensional Kondo lattice model, in agreement with previous works.
Based on the binding energy of two holes, we did not find evidence of superconductivity
in the 2D Kondo lattice model close to half-filling.
\end{abstract}


\maketitle
Recently, a great deal of excitement has been generated with the discovery of several
heavy fermion compounds which exhibit a quantum critical point (QCP) \cite{rstewart,pcol}.
Several anomalous behaviors have been observed close to the QCP and one of the main
problems of this area is to try to understand this puzzle. The Kondo lattice model
(KLM) is the simplest model believed to describe heavy fermion materials \cite{livro}.
This model incorporates the interaction between the localized spins and the conduction
electrons via an exchange interaction of strength $J$. This interaction is believed
to be the most important in heavy fermion compounds at low temperatures. At small
values of $J$, due to the indirect Ruderman-Kittel-Kasuya-Yosida (RKKY) interaction,
antiferromagnetic long-range order (LRO) is expected, whereas for large $J$ a paramagnetic
state emerges. The competition between these two states leads to the QCP, as first
pointed out by Doniach \cite{don}. Many theoretical works have concentrated on the
one-dimensional (1D) model, which can be treated more easily. In particular, at half-filling,
several numerical calculations show that the 1D KLM is always a gapped spin liquid
with no magnetic quasi-LRO \cite{1dkondo_qcp1,1dkondo_qcp2,rkondo}. Away from half-filling,
there is power law decay of spin correlations, for small values of $J$ (see, however,
Ref.~\onlinecite{dimer}). On the other hand, for the two-dimensional (2D) KLM athalf-filling,
 several studies support the existence of a QCP \cite{mtvqcp2,sexqcp2d,assaad}
(see also Ref. \onlinecite{pl}).

In this work, we present the spin and charge gaps of the N-leg Kondo ladders at half-filling
calculated by the density matrix renormalization group (DMRG) technique \cite{white}.
The N-leg Kondo ladders consist of $N$ Kondo chains of length $L$ coupled by the
hopping term. The 2D system is obtained by taking both $N$ and $L$ to infinity.
As far as we know this is the first study of the $N$-leg Kondo ladders. We will
also show evidence of the existence of a QCP in the 2D Kondo model at half-filling
in agreement with previous studies. Moreover, our results for the binding energy
of two holes do not show any evidence for an effective hole-hole attraction in the
2D KLM for any Kondo coupling. All the results we will report here correspond to
the conduction electron density $n=1$. It is important to note that for this density,
techniques other than the DMRG can be applied \cite{mtvqcp2,sexqcp2d,assaad}. However,
away from half-filling, methods like Quantum Monte Carlo\cite{assaad} are plagued
by the ``sign problem''. Therefore,  our results at half-filling should serve as benchmarks 
studies before get results at more general fillings. DMRG results away from half-filling are in progress and
will be presented elsewhere \cite{xavetal}.

We considered the N-leg Kondo Hamiltonian at half-filling with $M=N$x$L$ sites
\[
H=-\sum _{<i,j>,\sigma }(c_{i,\sigma }^{\dagger }c_{j,\sigma }^{\phantom {\dagger }}+\mathrm{H.}\, \mathrm{c.})+J\sum _{j}\mathbf{S}_{j}\cdot \mathbf{s}_{j}\]
 where $c_{j\sigma }$ annihilates a conduction electron in site $j$ with spin projection
$\sigma $, $\mathbf{S}_{j}$ is a localized spin $\frac{1}{2}$ operator, $\mathbf{s}_{j}=\frac{1}{2}\sum _{\alpha \beta }c_{j,\alpha }^{\dagger }\bm {\sigma }_{\alpha \beta }c_{j,\beta }^{\phantom {\dagger }}$
is the conduction electron spin density operator and $\bm {\sigma }_{\alpha \beta }$ are Pauli matrices. 
Here $<ij>$ denote near-neighbor
sites, $J>0$ is the Kondo coupling constant between the conduction electrons and
the local moments and the hopping amplitude was set to unity to fix the energy scale.
We investigated the model with the DMRG technique with open boundary conditions.
We used the finite-size algorithm for sizes up to $N\times L=40$, keeping up to
$m=2000$ states per block. The discarded weight was typically about $10^{-4}-10^{-8}$
in the final sweep. Although the DMRG is based on a one-dimensional algorithm, it
has been applied to low dimensional systems. The procedure consists in mapping the
low-dimensional model on a 1D model with long range interactions \cite{2d1,kago}.
Several important questions, such as the existence of striped phases and/or superconductivity
in the 2D $t-J$ model, have been addressed with this procedure \cite{tjrefs}. Here,
we used this same method to study the N-leg Kondo ladders.

Our first concern is to check the applicability of the technique to the N-leg Kondo
ladders. We can get some insight by looking at the ground state energies. We have
checked that, for small slightly doped clusters with up to 12 sites, we can reproduce
with the DMRG the same energies obtained independently by exact diagonalization (ED).
As we increase the system size the Hilbert space increases exponentially and ED is
no longer a suitable method to treat the system. This is the main advantage of the
DMRG. In order to illustrate this advantage, we will show that we can obtain the
ground state energies of the N-leg Kondo ladders in a controlled way \cite{comment}.

In table 1 we present the ground state energies as a function of the truncation $m$
for the 2-leg (4-leg) Kondo ladders with length $L=10$ $(L=4)$. The energies for
$m=\infty $ were obtained by an extrapolation assuming  a dependence of the type
$e(m)=e_{\infty }+a/m+b/m^{2}$. As we can see, for the $J=0$ (free case) with a
small number $m$ of states the relative error is smaller than $\sim 10^{-4}$. This
free case result does not appear to depend on the number of legs, as shown for the
2- and 4-leg ladders. However, a small coupling $J$ makes the wave function much
more complex and a larger number of states is necessary to keep the same accuracy.
Furthermore, in this weak coupling regime, we need a much larger number of states
(and consequently a longer computational time) to keep the algorithm stable as the
number of legs is increased. An increase in computer effort, probably exponential,
such as this has also been observed in other systems \cite{2d1,kago}. This was the
main reason why we restricted our calculations to $N\le 6$ and large values of $J$.
On the other hand, in the strong coupling limit, a smaller number of states is necessary
to keep the same accuracy. For $J\gg 0$ the conduction electrons form unbreakable
mobile singlets with the localized spins, effectively behaving like free spinless
fermions.

%
\scriptsize
\btb
\bc
\bt{|l|l|l||l|l|l||}                     \hline
             &  \mc{2}{c||}{2x10} & \mc{3}{c||}{4x4}          \\ \cline{3-5} \hline\hline
 $  _{\makebox[.4cm]{\large $m$}}    $  &$J$=0.0 & $J$=2.0 & $J$=0.0 & $J$=0.3 &$J=3.0$ \\ \cline{2-5} \hline\hline
100  & -27.56929  & -39.15411 & -21.85340 & -22.12936 & -41.11699  \\
 200  & -27.62807  & -39.15909 & -21.88557 & -22.19189 & -41.25039    \\
 300  & -27.63982  & -39.15988 & -21.88816 & -22.25239 & -41.28139 \\
 400  & -27.64290  & -39.16007 & -21.88845 & -22.29403 & -41.29605\\
 600  & -27.64407  & -39.16015 & -21.88853 & -22.33682 & -41.31698 \\
 800  &            &           &           & -22.36031 & -41.32227 \\
 1200  &           &           &           & -22.38726 & -41.32606 \\
 $\infty$& -27.6455 & -39.1602  & -21.8885   & -22.4413 & -41.3401
\\ \hline\hline
exact & -27.64486 & &-21.88854 &  &  \\  \hline
\et \ec \caption{The ground state energies of the Kondo laticce model for the clusters
 2x10 and 4x4  as a function of the number of states $m$ kept in the DMRG 
truncation for some
Kondo couplings. The energies at $m=\infty$ were obtained by an
extrapolation, see text. } \etb
%
\normalsize

\begin{figure}
\begin{center}\includegraphics[  width=2.8in,
  keepaspectratio]{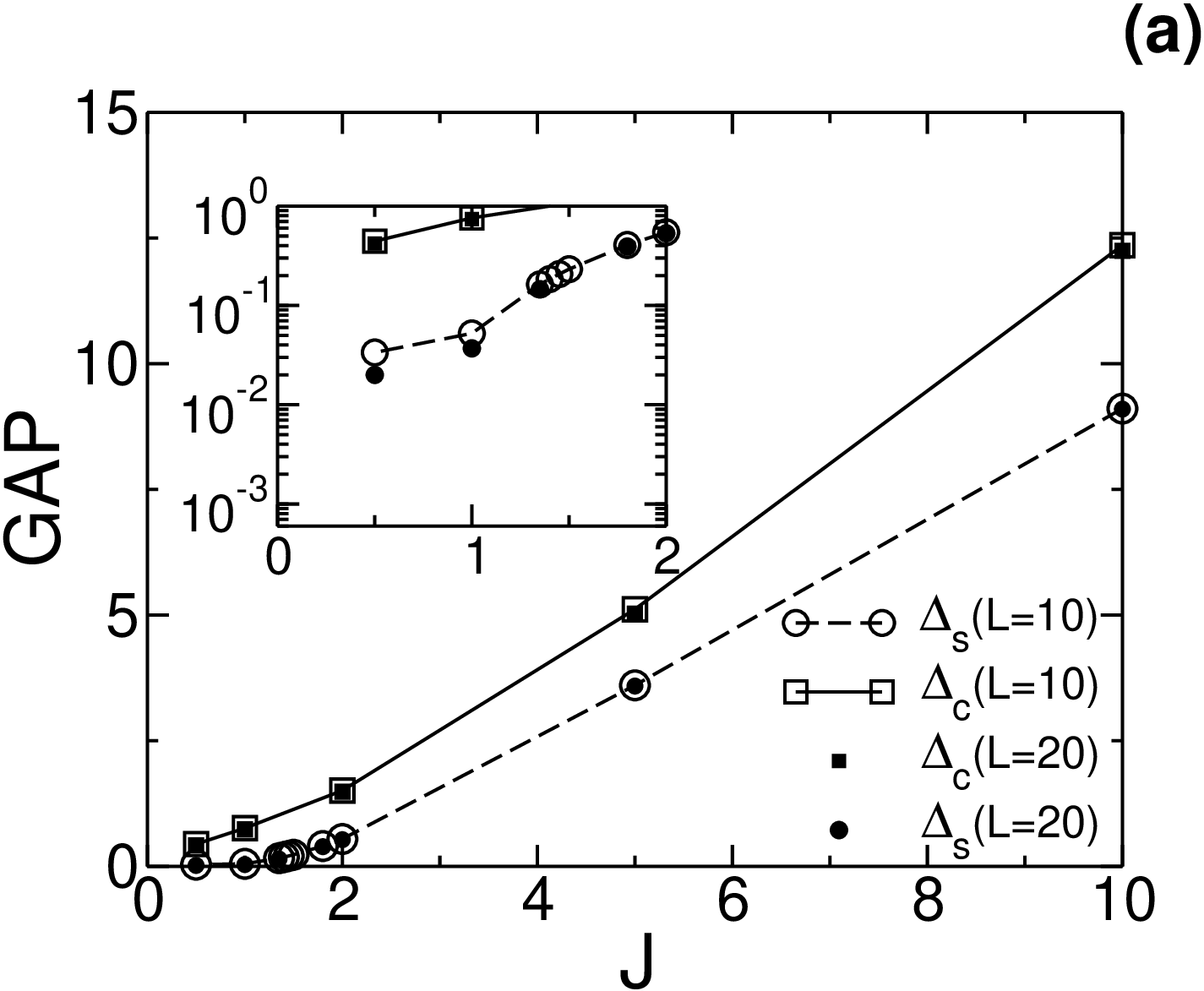}\end{center}

\begin{center}\includegraphics[  width=2.7in,
  keepaspectratio]{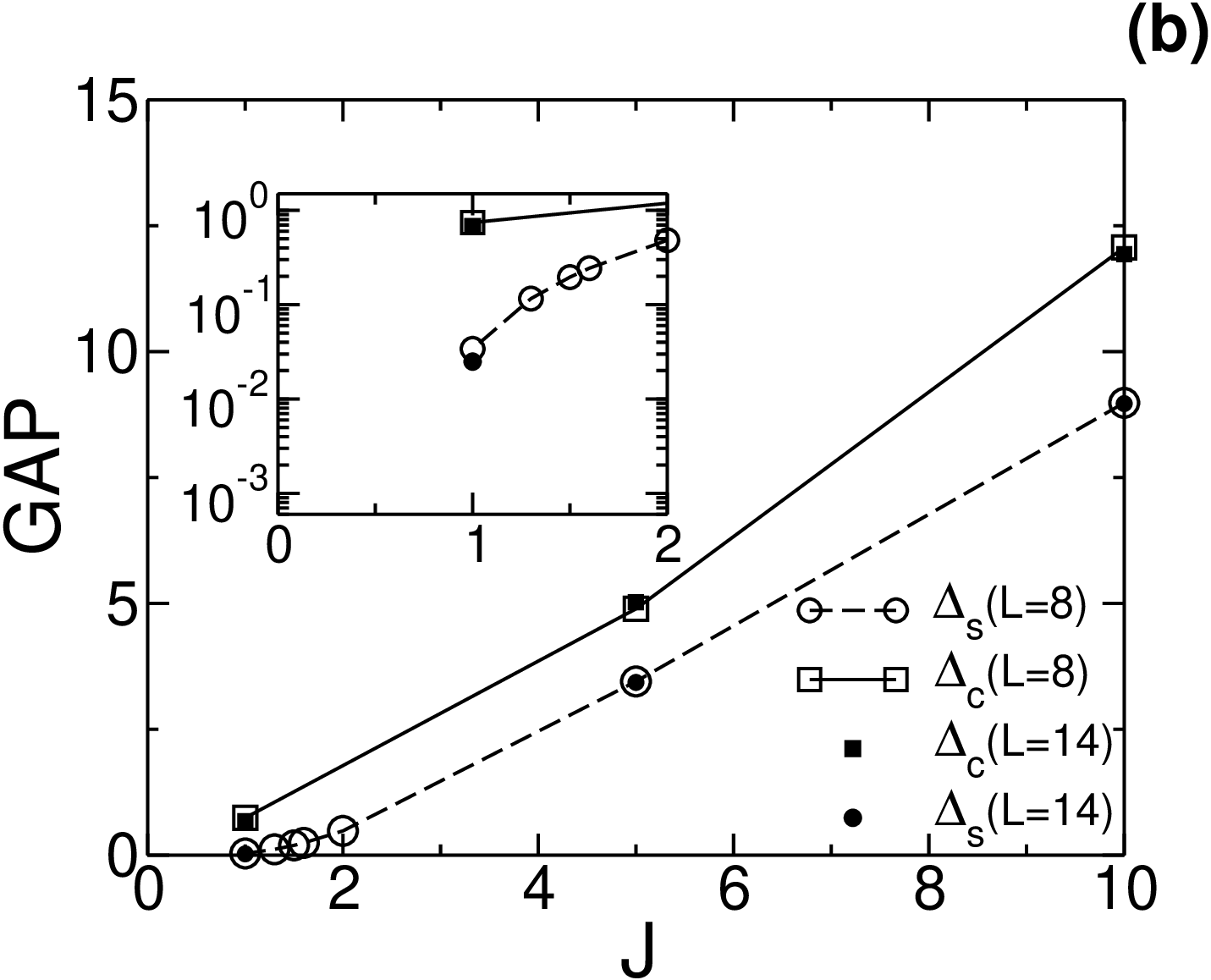}\end{center}

\caption{\label{fig1} (a) The spin and charge gaps for the 2-leg Kondo ladder with $L=10$
(open symbols) and $L=20$ (closed symbols). The circles (squares) correspond to
the spin (charge) gap. Inset: zoom of the region with $0<J<2$. (b) Same as (a),
but for the 3-leg Kondo ladder with $L=8$ (open symbols) and $L=14$ (closed symbols). }
\end{figure}

Having set these benchmarks, we are now able to calculate the spin and charge gaps
of the N-leg Kondo ladders with $M=N$x$L$ sites at half-filling. These are defined
as $\Delta _{s}(L)=e(M,1)-e(M,0)$ and $\Delta _{c}(L)=e(M-2,0)-e(0,0)$, respectively,
where $e(N_{c},s)$ is the ground state energy of the system with $N_{c}$ conduction
electrons in the subspace of total spin $s$ \cite{rkondo,spindec}. In Fig.~\ref{fig1}(a)
we present the spin and charge gaps for the 2-leg Kondo ladder with length $L=10$
(open symbols) and $L=20$ (closed symbols) as a function of the Kondo coupling $J$.
As we see, finite size effects are very small. In the strong-coupling limit the spin
and charge gaps increase linearly with $J$, and can be estimated by second order
perturbation theory with respect to $t/J$ \cite{rkondo}. At this order, the spin
and charge gaps of the $N$-leg Kondo ladders are given, respectively, 
by $\Delta _{s}^{(2)}=J-\frac{20\gamma ^{N}t^{2}}{3J}$
and $\Delta _{c}^{(2)}=3J/2-2\gamma ^{N}t+\gamma ^{N}(\frac{4\gamma ^{N}}{3}-1)\frac{t^{2}}{J},$
where $\gamma ^{N}=1-cos(\pi N/(N+1))$. We used the infinite-size algorithm
\cite{white} to determine the spin and charge gaps of the 2-leg Kondo ladders in
the strong coupling limit \cite{comment2}. Our numerical calculations are in very
good agreement with these results. For example, for $J=40$ we obtained $\Delta _{s}=39.752$
and $\Delta _{c}=57.057$, very close to the perturbative results $\Delta _{s}^{(2)}=39.750$
and $\Delta _{c}^{(2)}=57.037$. On the other hand, 
for small values of $J$ the spin gap has a different dependence
on the coupling constant.

The results we have found for the 2-leg Kondo ladder are very similar to those of
the 1D KLM\cite{rkondo}. In this case, for large $J$ the spin gap goes linearly
with $J$, while for small $J$ it decays exponentially as $\Delta _{s}\sim exp(-1/\alpha nJ)$,
where $\alpha \sim 1.4$ \cite{spindec}. This kind of behavior does not seem to
depend on the number of legs. The spin and charge gaps of the 3-leg Kondo ladder
show similar behavior, as shown in Fig.~\ref{fig1}(b). Based on the spin and charge
gap behaviors of the 1-, 2- and 3-leg Kondo ladders, we expect \textit{non-zero spin
and charge gaps} for the half-filled $N$-leg Kondo ladders. It is interesting to
note that the spin gap has similar behavior whether the number of legs is even or
odd. The same does not happen in $N$-leg Heisenberg ladders, where the spin gap
is zero for odd $N$ and different from zero for even $N$ \cite{2lelbio} (for a
review see Ref. \onlinecite{elbios}). 
This difference arises from the fact that 
the Kondo Ladders are coupled just with the 
hopping term, differently from the Heisenberg Ladders. So, the same analysis
that is  done in the N-leg Heisenberg ladders (see Ref. \onlinecite{elbios}), does
not apply in the N-leg Kondo ladders. Moreover, we might naively expect that
the inclusion of the Coulomb interaction $U$ in the N-leg Kondo ladders 
would  lead to a spin gap with a distinct dependence on  the number of legs (at least for $ U \gg 0$ ), 
as in the Heisenberg ladders. However, most probably, even the presence of $U$ will not lead to
this distinct behavior. Note that the 1D Kondo-Hubbard model at half-filling has a 
non-zero spin gap \cite{spindec}, differently 
from the 1D Heisenberg model.

We would still like to determine whether the spin gap remains finite or not when
$N$ goes to infinity (after $L\rightarrow \infty $). Approximate approaches \cite{mtvqcp2,sexqcp2d}
and Quantum Monte Carlo results \cite{assaad} support the existence of a QCP at
$J_{c}/t\sim 1.45\pm 0.05$. Below this value antiferromagnetic LRO appears with
a zero spin gap. It is a very tough task to determine the presence of LRO directly
by the value of the spin gap, since for small values of $J$ the gap is very small,
and as we increase the number of legs we lose accuracy in the energies.

\begin{figure}
\begin{center}\includegraphics[  width=3.5in,
  keepaspectratio]{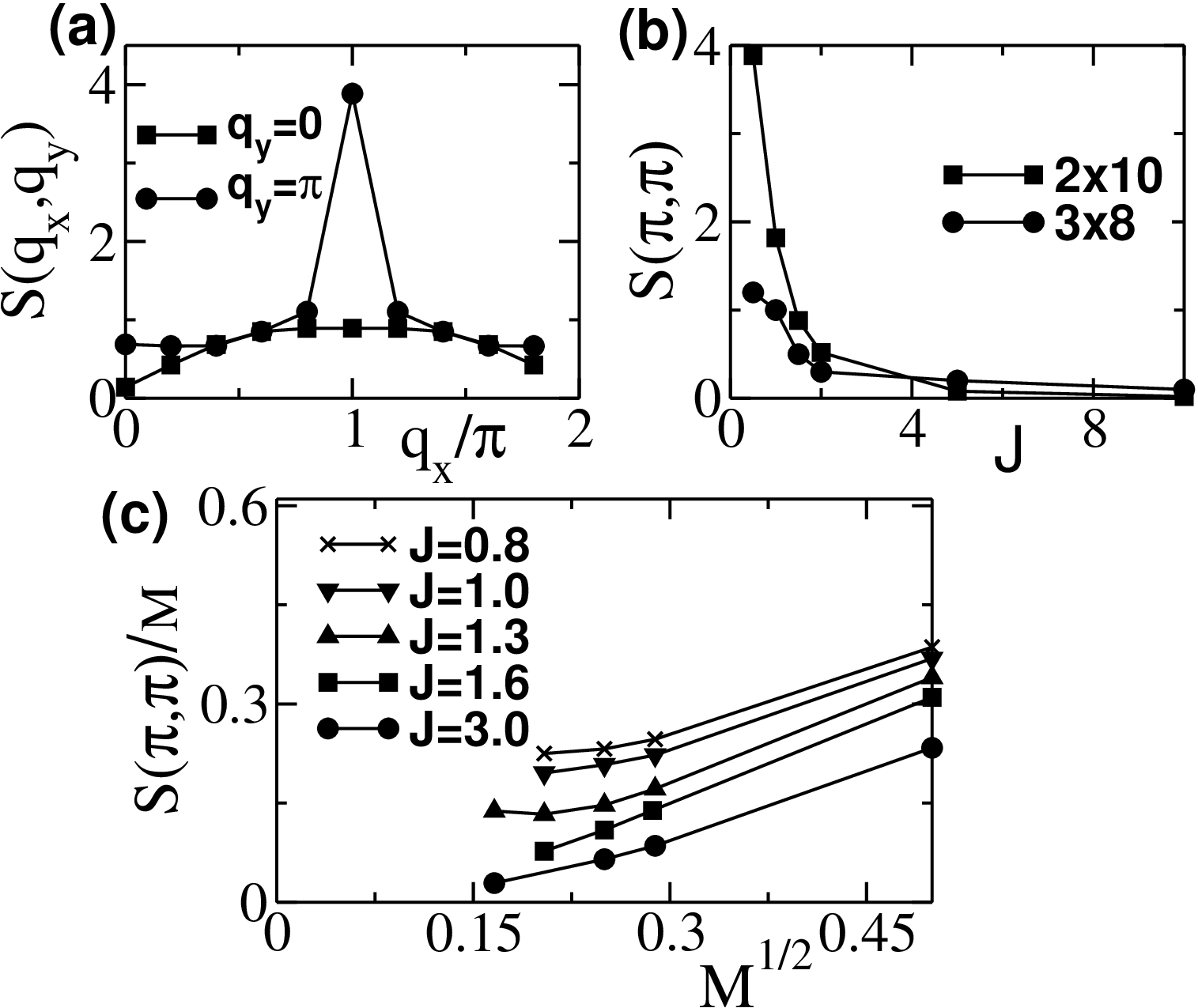}\end{center}

\caption{\label{fig2} The spin structure factor $S\left(\vec{q}\right)$ for the N-leg Kondo
ladders. (a) $S\left(\vec{q}\right)$ vs momentum $q_{x}$ for the 2-leg Kondo ladder
with length $L=10$ and $J=0.5$. (b) $S\left(\vec{q}=\left(\pi ,\pi \right)\right)$
vs the Kondo coupling $J$ for the clusters 2x10 (squares) and 3x8 (circles). (c)
$S\left(\vec{q}=\left(\pi ,\pi \right)\right)/M$ vs $1/M$ for several Kondo couplings.
The sizes of the clusters used were $M=$2x2, 3x4, 4x4, 4x6, and 6x6. }
\end{figure}

\begin{figure}
\begin{center}\includegraphics[  width=3.4in,
  keepaspectratio]{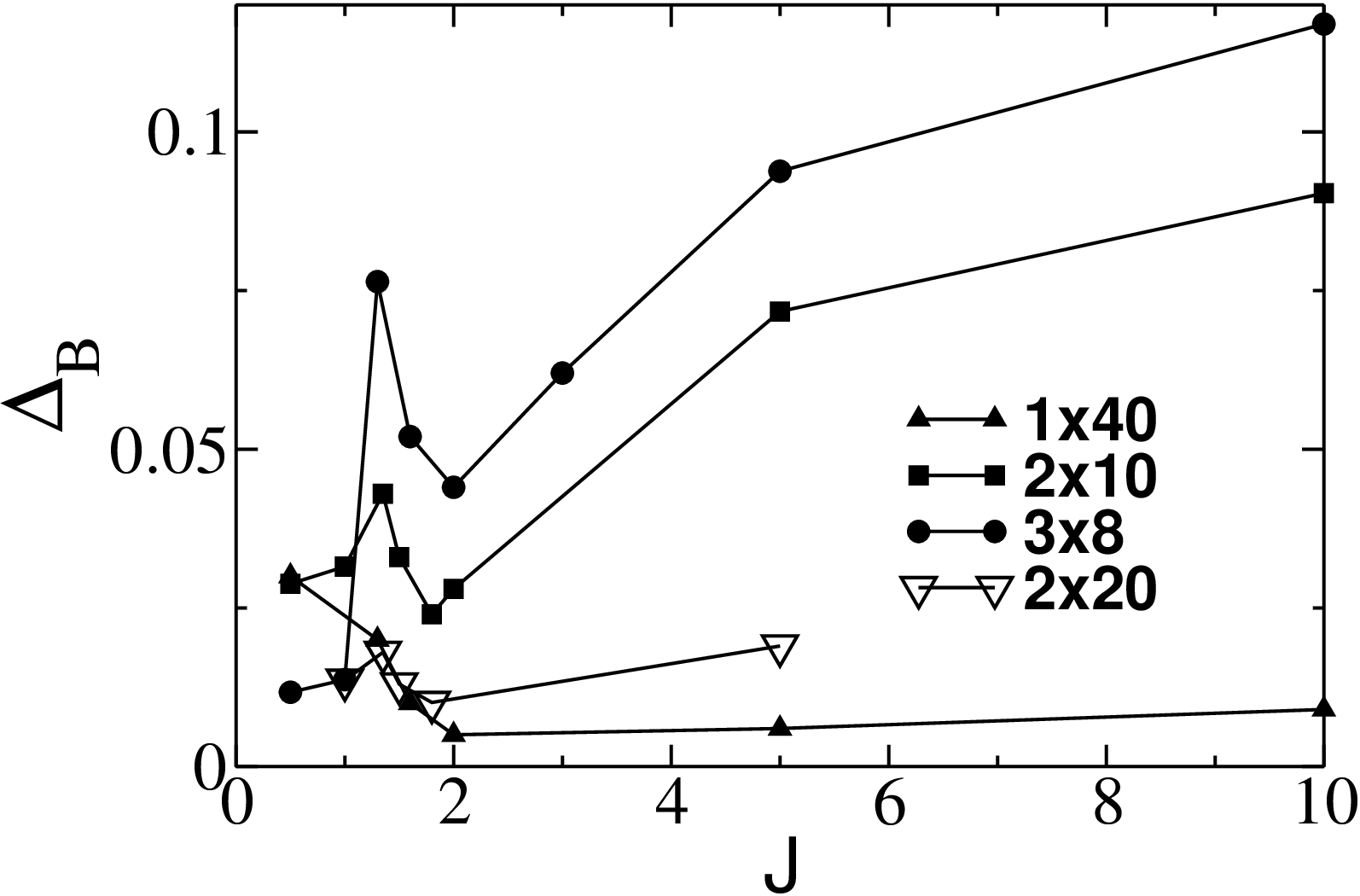}\end{center}

\caption{\label{fig3} The binding energy $\Delta _{B}$ for the 1-, 2- and 3-leg Kondo ladders
as a function of the Kondo coupling $J$. }
\end{figure}

Instead of looking at the spin gap to check for LRO, we can calculate the Fourier
transform of the spin-spin correlation function $S\left(\vec{q}\right)=\frac{1}{M}\sum _{\vec{r}_{1},\vec{r}_{2}}e^{\vec{q}\cdot \left(\vec{r}_{1}-\vec{r}_{2}\right)}\left\langle \mathbf{S}_{\vec{r}_{1}}\cdot \mathbf{S}_{\vec{r}_{2}}\right\rangle $.
In Fig.~\ref{fig2}(a), we show the spin structure factor $S\left(\vec{q}\right)$
for the 2-leg Kondo ladder as a function of momentum $q_{x}$ for $J=0.5$. The peak
at $\vec{q}=(\pi ,\pi )$ is a signature of some kind of antiferromagnetic order,
not necessarily long-ranged. For small values of $J$ we expect this peak to increase
as we decrease $J$, since the RKKY interaction starts to dominate over Kondo compensation.
Indeed, this is what we have seen. In Fig.~\ref{fig2}(b), we present the spin structure
factor $S\left(\vec{q}=\left(\pi ,\pi \right)\right)$, for the clusters 2x10 (squares)
and 3x8 (circles), as a function of the Kondo coupling $J$. Note the divergence
as $J\rightarrow 0$. This kind of behavior was also observed in the magnetic susceptibility
of the one-dimensional KLM \cite{1dkondo_qcp1}. Antiferromagnetic LRO exists if
$\lim _{M\to \infty }S\left(\vec{q}=(\pi ,\pi )\right)/M$ is non-zero. In Fig.~\ref{fig2}(c),
we present $S\left(\vec{q}=\left(\pi ,\pi \right)\right)/M$ as a function of $\sqrt{M}$
for several values of the Kondo coupling. As can be seen, even working with small
clusters, our results suggest that for $J>1.6$ $S\left(\vec{q}=(\pi ,\pi )\right)/M$
tends to zero as we increase $M$, while for $J<1.3$ it tends to a finite value.
This result is compatible with $J_{c}=1.45$ as suggested by other methods \cite{mtvqcp2,sexqcp2d,assaad}.

Finally, we would like to check for evidence of superconductivity in the 2D KLM close
to the QCP. Several heavy fermions compounds, like CeIn$_{3}$ and CePd$_{2}$Si$_{2}$,
can be tuned to an antiferromagnetic QCP by application of external pressure \cite{exp1}.
Superconductivity was observed at low temperatures close to the QCP of these materials.
Our aim will not be to find proof of superconductivity at finite hole doping close
to a quantum critical point in the 2D KLM, which would entail the calculation of
the pair correlation function. Instead, we can gain some insight by looking at the
binding energy of two doped holes, defined as $\Delta _{B}=E(2)+E(0)-2E(1)$, where
$E(n)$ is the ground state energy with $n$ holes. If the holes are bound $\Delta _{B}<0$,
and this is indicative that attractive effective forces are present \cite{relbio}.
This is a necessary, though not sufficient, condition for a possible condensation
of the holes. In Fig.~\ref{fig3}, we present the hole binding energy for the 1-,
2- and 3-leg Kondo ladders. In the  thermodynamic limit the binding energies vanish if holes
do not form bound states and is asymptotically positive \cite{relbio}.
As we see from this figure, this is exactly what we have observed.
Increasing the cluster from 2x10 to 2x20  the binding energies
decrease. As we see, the holes do not bind for \emph{any} Kondo
coupling. Similar results were also observed in the Kondo chain \cite{1dkondo_qcp1}.
These results suggest that, at least close to half-filling, superconductivity will
not emerge in the 2D Kondo lattice model. 
The fact that we  have not found bind holes  also suggest
the the KLM may  not be the minimum model to describe correctly the Heavy
Fermion compounds.

In summary, we presented, for the first time, a systematic study of the 2- and 3-leg
Kondo ladders at half-filling by using the DMRG. Based on the behavior of the spin
gap of the 1- to 3-leg Kondo ladders, we expect non-zero charge and spin gaps in
the N-leg Kondo ladders for any $J>0$. Using small clusters we were able to determine
the presence of a QCP, in agreement with previous works. Our results also suggest
that there is no superconductivity close to half-filling in the 2D Kondo lattice
model.

The author thanks E. Miranda, E. Dagotto, P. G. Pagliuso, A. H. Castro-Neto, and
R. R. dos Santos for useful discussions. This work was supported by FAPESP 00/02802-7
and 01/00719-8.


\begin{thebibliography}{10}
\bibitem{comment}
In the DMRG it is believed that the energy error is proportional to the truncation error \cite{white}. 
However, at the present moment, there is no exact relation between them. A possible way to
estimate the relative error  is the following.  We define the  relative error as 
$\left(e\left(m=\infty\right)-e\left(m_{last}\right)\right)/e\left(m=\infty\right)$ ,
where $e\left(m=\infty\right)$ is the extrapolated energy (see text) and $e\left(m_{last}\right)$ 
is the energy with the biggest $m$ available. In all results presented here, the relative error is 
smaller than $0.4\times10^{-4}$, $0.2\times10^{-3}$, $0.3\times10^{-3}$ and $0.7\times10^{-3}$ 
for the 2-, 3-, 4-  and 6-leg Kondo ladders respectively.
\bibitem{comment2}The infinite-size algorithm is very stable for large $J$. For small values of $J$ we
were not able to get precise results using the same procedure.
\end{thebibliography}
\end{document}